\documentclass[12pt]{iopart}             % Journal Style
\usepackage{iopams}
\usepackage{graphicx}
\usepackage{url,graphics,epsfig}
\usepackage{amssymb}
\usepackage[breaklinks=true,colorlinks=true,linkcolor=blue,urlcolor=blue,citecolor=blue]{hyperref}
\usepackage{cite}
\usepackage{dcolumn}

\begin{document}
\jl{2}
%
%+++++++++++++++++++++++++++++++++++++++++++++++++++++++++++++++++++++++++++++
%
%  Macro definitions
%
%+++++++++++++++++++++++++++++++++++++++++++++++++++++++++++++++++++++++++++++
\def\etal{{\it et al~}}
\def\newblock{\hskip .11em plus .33em minus .07em}
%+++++++++++++++++++++++++++++++++++++++++++++++++++++++++++++++++++++++++++++
%  End of Macro definitions
%
%+++++++++++++++++++++++++++++++++++++++++++++++++++++++++++++++++++++++++++++
%
%+++++++++++++++++++++++++++++++++++++++++++++++++++++++++++++++++++++++++++++
%
% Title of the paper
%
%+++++++++++++++++++++++++++++++++++++++++++++++++++++++++++++++++++++++++++++
%
\setlength{\arraycolsep}{2.5pt}             % use this for journal style

\title[Valence shell photoionization of  W$^{5+}$ ions] {Photoionization of tungsten ions:\\  experiment  and theory for W$^{5+}$}

\author{A M\"{u}ller$^1\footnote[1]{Corresponding author, E-mail: Alfred.Mueller@iamp.physik.uni-giessen.de}$,
		S Schippers$^{1,2}$, J Hellhund$^{1,4}$, A L D Kilcoyne$^3$,\\ R A Phaneuf$^4$
	           and B M McLaughlin$^{5,6}\footnote[2]{Corresponding author, E-mail: bmclaughlin899@btinternet.com}$}

\address{$^1$Institut f\"{u}r Atom- und Molek\"{u}lphysik,
                         Justus-Liebig-Universit\"{a}t Gie{\ss}en, 35392 Giessen, Germany}

\address{$^2$I. Physikalisches Institut,
                           Justus-Liebig-Universit\"{a}t Gie{\ss}en, 35392 Giessen, Germany}

\address{$^3$Advanced Light Source, Lawrence Berkeley National Laboratory,
                          Berkeley, California 94720, USA }

\address{$^4$Department of Physics, University of Nevada,
                          Reno, NV 89557, USA}

\address{$^5$Centre for Theoretical Atomic, Molecular and Optical Physics (CTAMOP),
                          School of Mathematics and Physics,
                          Queen's University Belfast, Belfast BT7 1NN, UK}

\address{$^6$Institute for Theoretical Atomic and Molecular Physics,
                          Harvard Smithsonian Center for Astrophysics, MS-14,
                          Cambridge, MA 02138, USA}
%
%+++++++++++++++++++++++++++++++++++++++++++++++++++++++++++++++++++++++++++++
%
%              Abstract
%
%+++++++++++++++++++++++++++++++++++++++++++++++++++++++++++++++++++++++++++++

\begin{abstract}
Experimental and theoretical cross sections are reported  for single-photon single ionization of  W$^{5+}$ ions. Absolute measurements were conducted   employing the photon-ion merged-beams technique. Detailed photon-energy scans were performed  at (67$\pm$10)~meV resolution in the 20 -- 160 eV range. In contrast to photoionization of tungsten ions in lower charge states, the cross section is dominated by narrow, densely-spaced resonances.  Theoretical results were obtained from a Dirac-Coulomb R-matrix approach employing a basis set of 457 levels providing cross sections  for photoionization of W$^{5+}$ ions in the $4f^{14}5s^2 5p^6 5d \; {^2}{\rm D}_{3/2}$ ground level as well as the $4f^{14}5s^2 5p^6 5d \; {^2}{\rm D}_{5/2}$ and  $4f^{14}5s^2 5p^6 6s \; {^2}{\rm S}_{1/2}$ metastable excited levels.
Considering the complexity of the electronic structure of tungsten ions in low charge states, the agreement between theory and experiment is satisfactory.
\end{abstract}

%\pacs{32.80.Fb, 31.15.Ar, 32.80.Hd, and 32.70.-n}

\vspace{0.25cm}
\begin{flushleft}
Short title: Valence shell photoionization of W$^{5+}$ ions\\
\vspace{0.25cm}
\submitto{\jpb: \today, Draft }
\end{flushleft}
\maketitle
%      Text of paper follows
%
%++++++++++++++++++++++++++++++++++++++++++++++++++++++++++++++++++++++++++++
\section{Introduction}
Tungsten and its ions in low charge states are prototypical for heavy atoms with a complex electronic structure. While level energies and cross sections for fundamental processes of light few-electron systems are fairly well understood, computations for many-electron systems are limited in accuracy by the complexity of the underlying physics and the limitations in the available computing resources. Progress towards improved prediction and description  of the structure and the dynamics of complex atoms is highly desirable for fundamental and application-related reasons.

A very important application of tungsten is in its envisaged use in controlled-nuclear-fusion reactors. Its unique physical and chemical properties make it the most suitable material for the wall regions of highest particle and heat load in a fusion reactor vessel~\cite{Neu2013}. The downside of tungsten as an impurity is its extremely high potential for radiative plasma cooling~\cite{Neu2003,Puetterich2010a}. For modeling and controlling the plasma it is essential to understand the collisional and spectroscopic properties of tungsten atoms and ions.  In order to meet some of the most important data requirements, dedicated experimental and theoretical projects have been initiated with the goal to provide cross section data and spectroscopic information on tungsten ions interacting with electrons and photons~\cite{Mueller2015b,Beiersdorfer2015,Preval2017}.  The present work provides experimental and theoretical cross sections for single-photon single ionization of W$^{5+}$ ions. It completes a series of photoionization studies on tungsten atoms~\cite{Ballance2015a} and ions in low charge states~\cite{Mueller2011a,Mueller2012,Mueller2014c,Mueller2015h,McLaughlin2016a,Mueller2017b}. It is noted that radiative properties and core-polarization effects in the W$^{5+}$ ion have been investigated  by Enzonga Yoca \etal~\cite{EnzongaYoca2012a}.

Photoionization of tungsten atoms and ions, although not of direct relevance to fusion research, is of plasma-related interest nevertheless because
it can provide details about spectroscopic properties of tungsten which are needed for plasma diagnostics. The present Dirac-Coulomb R-matrix approximation is one of the most advanced theoretical tools to generate data on electron-ion and photon-ion interactions. Studies on photoionization of tungsten atoms and ions with their complex electronic structure featuring open $d$ and $f$ shells  and comparison of experimental and theoretical results can provide benchmarks and guidance for future theoretical work on electron-ion and photon-ion interaction processes of complex many-electron systems.

In this paper we report on experimental and theoretical cross sections for single photoionization of  W$^{5+}$ ions.  The layout of this paper is  as follows. Section 2 details the experimental procedure.
Section 3 presents a brief outline of the theoretical work. Section 4 presents a discussion of the
results obtained from both the experimental and theoretical methods.
Section~\ref{sec:Conclusions} summarizes the results and draws conclusions from the present investigation.
%
%##########################################################################################
%
%       Experimental section of the paper
%
%##########################################################################################
%
%

%##########################################################################################
%
%       Fig 0
%
%##########################################################################################
%
\begin{figure*}
\centering
\includegraphics[width=17cm]{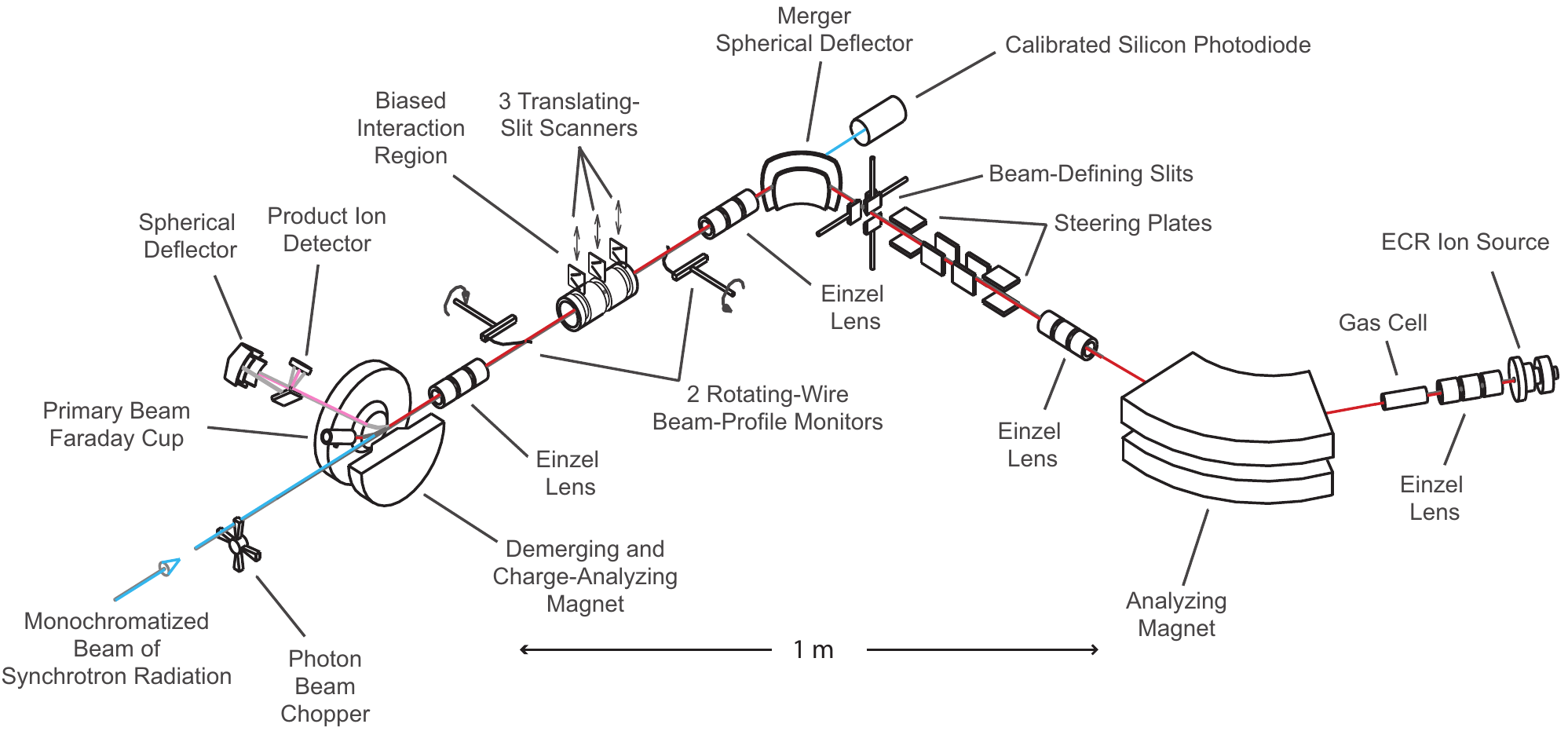}
\caption{\label{Fig:IPB}(Colour online)
Schematic sketch of the experimental setup~\cite{Phaneuf2007}.
}
\end{figure*}

\section{Experiment}\label{sec:exp}

For the experiments on photoionization of  W$^{5+}$  ions the Ion-Photon Beam (IPB) endstation (see Fig.~\ref{Fig:IPB}) of beamline 10.0.1.2 at the Advanced Light Source (ALS) in Berkeley was employed. The measurement of absolute cross sections made use of the merged-beams technique~\cite{Schippers2016}. The general layout of the  IPB setup and the associated experimental procedures were previously documented by Covington \etal~\cite{Covington2002a}. Since then significant technological improvements have been implemented.  A detailed description of the methodology used for the photoionization of tungsten ions has been provided in our publication on the results for W$^+$ ions~\cite{Mueller2015h} and most recently in connection with our results on W$^{4+}$ ions~\cite{Mueller2017b}.

Here, only an overview of the experiment is presented with specific aspects of the present measurements illuminated in more detail. The experimental arrangement is sketched in Fig.~\ref{Fig:IPB}. Tungsten ions were produced by leaking W(CO)$_6$ vapour via a needle valve into the plasma chamber of a 10~GHz electron-cyclotron-resonance (ECR) ion source. A discharge was maintained by adding Ar or Xe as a support gas.  The ions produced in the source were extracted and  accelerated by a voltage of $U_{\rm acc}$ = 6~kV. An ion beam was formed by a suitable set of electrostatic focusing elements. By a subsequent 60$^o$ dipole magnet the ion beam components were dispersed according to their charge and mass. The desired $^{186}$W$^{5+}$ beam component was selected and transported to an electrostatic spherical 90$^o$ deflector (the merger) which directed the ion beam onto the photon beam axis.  Ion currents of collimated beams of isotope-resolved W$^{5+}$ ions employed in the present experiments reached 58~nA.

Beyond the merger the selected W$^{5+}$ ion beam passed the  interaction region which was defined by an electrically isolated drift tube of  29.4~cm length. For the measurement of absolute cross sections the interaction region was set to a potential of up to $U_{\rm D}$ = 1~kV in order to tag product ions  with electrical charge $6e$ arising from within the interaction region by their final energy $5e U_{\rm acc} + e U_{\rm D}$. A 45$^o$ dipole magnet, the demerger, separated the W$^{5+}$ parent ion beam from the W$^{6+}$ products which were further deflected out of plane by a spherical 90$^o$ deflector and directed toward a single-particle detector with close to 100\% efficiency. The energy difference $e U_{\rm D}$ between product ions from outside and from inside the interaction region was sufficient for complete separation of the two components by the demerger magnet. The primary ion beam current was collected by a large Faraday cup inside the demerger magnet. Separation of photoionized ions from background  was accomplished by mechanically chopping the photon beam and by phase-sensitive recording of detector pulses.

For absolute cross section measurements the beam overlap factor~\cite{Schippers2016} was determined by scanning x- and y-profiles of the ion beam and the counter-propagating photon beam  at three positions in the middle and at the front and rear ends of the interaction region. By energy-tagging the product ions the length of the interaction region was defined as the length of the isolated drift tube  needed for absolute cross-section determinations. The error budget of the absolute cross sections obtained by this procedure has been discussed previously~\cite{Mueller2014b} and a total systematic uncertainty of 19\% was estimated. This uncertainty does not address problems with the purity of the two merging beams. In a thorough investigation described by M\"{u}ller \etal in the context of photoionization of W$^+$ ions~\cite{Mueller2015h}, energy-dependent fractions of higher-order radiation in the photon beam up to the sixth order could be detected and quantified. A procedure for correcting measured apparent cross sections was developed both with respect to correct normalization of photoionization signal to the photon flux and removal of surplus signal arising from photoionization at the higher energies $nE_\gamma$ of the $n$th order radiation fractions. We note that even-order contributions could be only partly suppressed by tightly closing the baffles downstream of the monochromator and thereby losing most of the flux of the photon beam. Corrections were made for the second- and third-order contaminations of the photon beam, neglecting the smaller effects of radiation orders $n \geq 4$. The uncertainties of this procedure were added to the total possible error of the measured absolute cross sections as described previously~\cite{Mueller2015h}. The uncertainty of the energy scale in the present experiments is estimated to be $\pm 200$~meV.

A further problem in experiments employing beams of ions with a complex electronic structure is the possible presence of ions in long-lived excited states. Obvious candidates for metastable excited states in W$^{5+}$ are associated with the $5p^65d$ ground-state and equal-parity $5p^66s$ configurations.  W$^{5+}$ has a $5p^65d~^2{\rm D}_{3/2}$ ground level, with an ionization potential of 64.77 $\pm$ 0.4~eV~\cite{NIST2019}. The least-excited metastable levels  are $5p^65d~^2{\rm D}_{5/2}$ and $5p^66s~^2{\rm S}_{1/2}$ with ionization potentials of 63.69 and 54.92~eV, respectively. In addition to these outer-shell excited levels, states within inner-valence-shell excited configurations, such as $5p^5 5d^2$, $4f^{13} 5d^2$, and $4f^{13} 5d 6s$ with ionization potentials less than about 31~eV, must be considered.

In separate experiments, cross sections and detailed energy-scan spectra have been measured for electron-impact ionization of W$^{5+}$ ions~\cite{Jonauskas2019a} employing an  ECR ion source identical to the one used in the present photoionization experiment. On the basis of detailed calculations of cross sections for electron-impact ionization of W$^{5+}$ ions in different long-lived levels, their fractional abundance in the parent ion beam used for the measurements was analyzed~\cite{Jonauskas2019a}. The analysis indicated that (85$\pm$9)\% of the ions in the primary beam were in the ground configuration. Since the ion source for the present experiment was operated similarly to optimize W$^{5+}$ production, the ion beam compositions are expected to be similar. Comparison of the present experimental results with the present calculations suggest a 2.5\% fraction of W$^{5+}(5p^6 6s~^2{\rm S}_{1/2})$ ions in the parent ion beam employed for the measurement of photoionization cross sections (see Sec.~\ref{sec:Results}).

%
%##########################################################################################
%
%       Theory section of the paper
%
%##########################################################################################
%
%
\section{Theory}\label{sec:Theory}

 The present work employs an efficient parallel version~\cite{Ballance2006,Fivet2012}  of the
 Dirac-Atomic {\textit R}-matrix-Codes (DARC)
 \cite{Norrington1987,Wijesundera1991,darc,McLaughlin2012a,McLaughlin2012b} developed for treating electron and photon interactions with atomic systems.  This suite continues to evolve~(\cite{McLaughlin2019a} and references therein)
 %\cite{McLaughlin2015a, McLaughlin2015b, McLaughlin2016c, McLaughlin2017b, McLaughlin2018, McLaughlin2019}
in order to address ever  increasing expansions for the target
and collision models used in electron and photon impact with heavy
atomic systems.

For a quantitative understanding of the experimental results on photoionization  of W$^{5+}$ ions, theory faces a number of problems which are much less pronounced for light, few-electron atoms and ions.

One difficulty has already been addressed in Sec.~\ref{sec:exp}, namely, the possible presence of ions in long-lived excited levels in the W$^{5+}$ parent ion beam that was used in the experiments. While such metastable levels can also be populated in light, few-electron systems, their number is particularly large in complex ions such as W$^{5+}$. Beside the $5p^6 5d~^2{\rm D}_{3/2}$ ground level, the excited fine-structure level $5p^6 5d~^2{\rm D}_{5/2}$ of the ground configuration and the lowest excited configuration $5p^6 6s$ with the level $^2{\rm S}_{1/2}$ contribute to the measured signal. Moreover, contributions from levels in the $5p^5 5d^2$ and $4f^{13} 5p^6 5d^2$ inner-subshell-excited configurations have to be considered as demonstrated by Jonauskas \etal~\cite{Jonauskas2019a} who found a negligibly small contribution ($\approx$0.7\%) of levels in the $4f^{13} 5p^6 5d 6s$ parent-ion configuration. The number of levels in the remaining two initial configurations amounts to $45 + 81 = 126$, a number that is prohibitively large for  DARC calculations of photoionization of a complex multi-electron ion such as W$^{5+}$. Even at the most powerful supercomputer facilities used world wide, the present calculations would exceed the limitations of presently available computing resources.  These contributions from  126 levels on top of the 3 initial levels considered, were neglected in the present theoretical modeling of the experimental data.

A further difficulty originates from the relatively close energy-spacing between the outermost $4f$, $5s$, $5p$, and $5p$ subshells in the parent ions and an even more dense multitude of autoionizing levels that can be excited  by the incident photons. The associated configuration interactions and correlation effects require very large basis sets for a suitable description of the electronic structures and the photoionization cross sections. Again, the presently available computing resources put serious limits on the number of configurations and levels that can be integrated into the basis set used in the DARC calculations. In order to provide a suitable representation of experimental results with the given constraints, an optimized basis set has to be constructed that contains the most important configurations and associated levels and thus, although being finite, gives the best possible representation of the physics behind the investigated problem within the limitations of the computing resources. Although hundreds of levels of the W$^{6+}$ target ion are included in the present treatment, the accuracy of the calculated resonance features (resonance energy and strength) cannot be expected to be completely satisfactory.

\subsection{Electronic structure}
\label{subsec:structure}
The investigation of single photoionization of the  W$^{5+}$ ion began with a
simple 60-level  approximation for the residual W$^{6+}$ target wavefunctions arising from the 7 configurations
$4f^{14}5s^25p^6$, $4f^{14}5s5p^65d$,  $4f^{14}5s^25p^55d$, $4f^{14}5p^65d^2$,
 $4f^{14}5s^25p^56s$,  $4f^{14}5s^25p^56p$, and $4f^{13}5s^25p^65d$.
To this a further set of 4 configurations;  $4f^{14}5s^25p^56d$,
 $4f^{14}5s^25p^57s$,  $4f^{14}5s^25p^57p$, and  $4f^{14}5s^25p^57d$ was added, resulting in a 98-level model for the residual target wavefunctions.

In order to explore  electron correlation further, the 60-level model was extended by  providing  for one- and selective two-electron
promotions to the $5d$, $6s$, $6p$ and $6d$ orbitals.  This included one-electron promotions from
the $4f$-shell to the $5d$, $6s$, $6p$ and $6d$ orbitals, giving the additional configurations $4f^{13}5s^25p^66s$, $4f^{13}5s^25p^66p$, and $4f^{13}5s^25p^66d$. We also added selective  two-electron promotions, namely, $4f^{14}5s^25p^45d^2$, $4f^{14}5s^25p^46s^2$,  $4f^{14}5s^25p^46p^2$,  $4f^{14}5s^25p^46d^2$, $4f^{14}5p^66s^2$,  $4f^{14}5p^66p^2$,  $4f^{14}5p^66d^2$, and $4f^{14}5s^25p^45d6s$. This gave us 19 configuration-state functions for our model and a total of 457 levels forming the basis for describing the residual W$^{6+}$  ion.

%
% Table 1
%
\begin{table}
%\begin{flushleft}
\small
\caption{Comparison of the NIST~\cite{NIST2019} tabulated data with the present theoretical energies
	   obtained by using the GRASP code~\cite{Dyall1989}. Relative energies with respect to the ground state are
             given  in Rydbergs.  A sample of the  lowest nine NIST levels of the residual W$^{6+}$ ion are
             compared with various level GRASP calculations.}
\label{tab1}.
%\begin{indented}
\begin{tabular}{llcccccc}
\br
Level         	&  Config 	     		&  Term		&$J$	& NIST  		 	&GRASP		&GRASP 	    	& GRASP	\\
		&				&			&	& Energy$^{\dagger}$ 	&Energy$^a$	&Energy$^b$	& Energy$^c$  \\	
		&				&			&	& (Ry)		 		&(Ry)			&(Ry) 			& (Ry)   \\	
\mr
 1  		& $4f^{14}5s^25p^6$ 	&  $\rm^1S$  	&0   	& 0.00000			&0.00000		&0.00000		&0.00000 \\
\\
 2  		& $4f^{13}5s^25p^65d$ &  $(7/2,3/2)^o$	&2     	& 2.83806			&2.82664		&2.94975		&2.98381 \\
 3  		&				&  $(7/2,3/2)^o$	&5     	& 2.89952			&2.90287		&3.02561		&3.06043 \\
 4  		& 			 	&  $(7/2,3/2)^o$	&3     	& 2.92237			&2.92561		&3.04983		&3.08525 \\
 5  		& 			 	&  $(7/2,3/2)^o$	&4     	& 2.93908			&2.94311		&3.06837		&3.10409 \\
\\
 6  		& $4f^{14}5s^25p^55d$ &  $(3/2,3/2)^o$	&0     	& 2.87285			& 2.82441		&2.95654		&2.99351 \\
 7  		& 			 	&  $(3/2,3/2)^o$	&1     	& 2.90602			&2.86927		&2.99914		&3.03556 \\
 8  		& 			 	&  $(3/2,3/2)^o$	&2     	& 3.05290			&2.99162		&3.11510		&3.15063 \\
 9  		& 				&  $(3/2,3/2)^o$	&3     	& 3.05431			&3.02244		&3.15012		&3.18691 \\
\mr
\end{tabular}
\\
\begin{flushleft}
$^{\dagger}$Energies from the NIST Atomic Spectra Database  \cite{NIST2019}.\\
$^a$GRASP theoretical energies from the 60-level approximation.\\
$^b$GRASP theoretical energies from the 98-level approximation.\\
$^c$GRASP theoretical energies from the 457-level approximation.\\
%\end{indented}
\end{flushleft}
\end{table}
Table~\ref{tab1} compares the excitation energies of the lowest  W$^{6+}$ levels listed in the NIST Atomic Spectra Database~\cite{NIST2019} with the results obtained  from three different approximations used to describe the residual  W$^{6+}$  ion by employing the GRASP code~\cite{Dyall1989}. As can clearly be seen from Table~\ref{tab1} the energy levels are only slowly converging as  noted in our previous studies on low charge states of tungsten ions \cite{Mueller2015h,McLaughlin2016a,Mueller2017b}.  Even with the largest calculation for the 457-level approximation, the reference energies provided by the NIST Atomic Spectra Database~\cite{NIST2019} are only approximately reproduced by the GRASP calculations, with discrepancies of up to about 2.2~eV. Given the limitations in existing computing resources this present approximation cannot be substantially extended and improved with respect to an appropriate target representation for subsequent scattering models.

\subsection{Photoionization calculations}
For the $5p^6 5d~^2{\rm D}_{3/2}$ ground level, as well as the $5p^6 5d~^2{\rm D}_{5/2}$ and $5p^6 6s~^2{\rm S}_{1/2}$ metastable initial levels of the tungsten ions studied here, the outer region electron-ion problem was
solved (in the resonance region below and  between all thresholds) using a fine energy mesh with steps of $\approx$ 24.5 $\mu$eV, and $\approx$ 245 $\mu$eV in the region where no pronounced resonances are observed.
The $jj$-coupled Hamiltonian diagonal matrices were adjusted so that the theoretical term energies matched the recommended
NIST values~\cite{NIST2019}. This is meant to improve the positioning of resonances relative to all thresholds included in the calculations. The theoretical DARC photoionization  cross sections from the 457-level calculations were convoluted with a 67~meV full-width-at-half-maximum (FWHM) Gaussian function in order to simulate the experimental photon energy resolution.

\section{Results and Discussion}\label{sec:Results}

\begin{figure*}
\centering
\includegraphics[width=17cm]{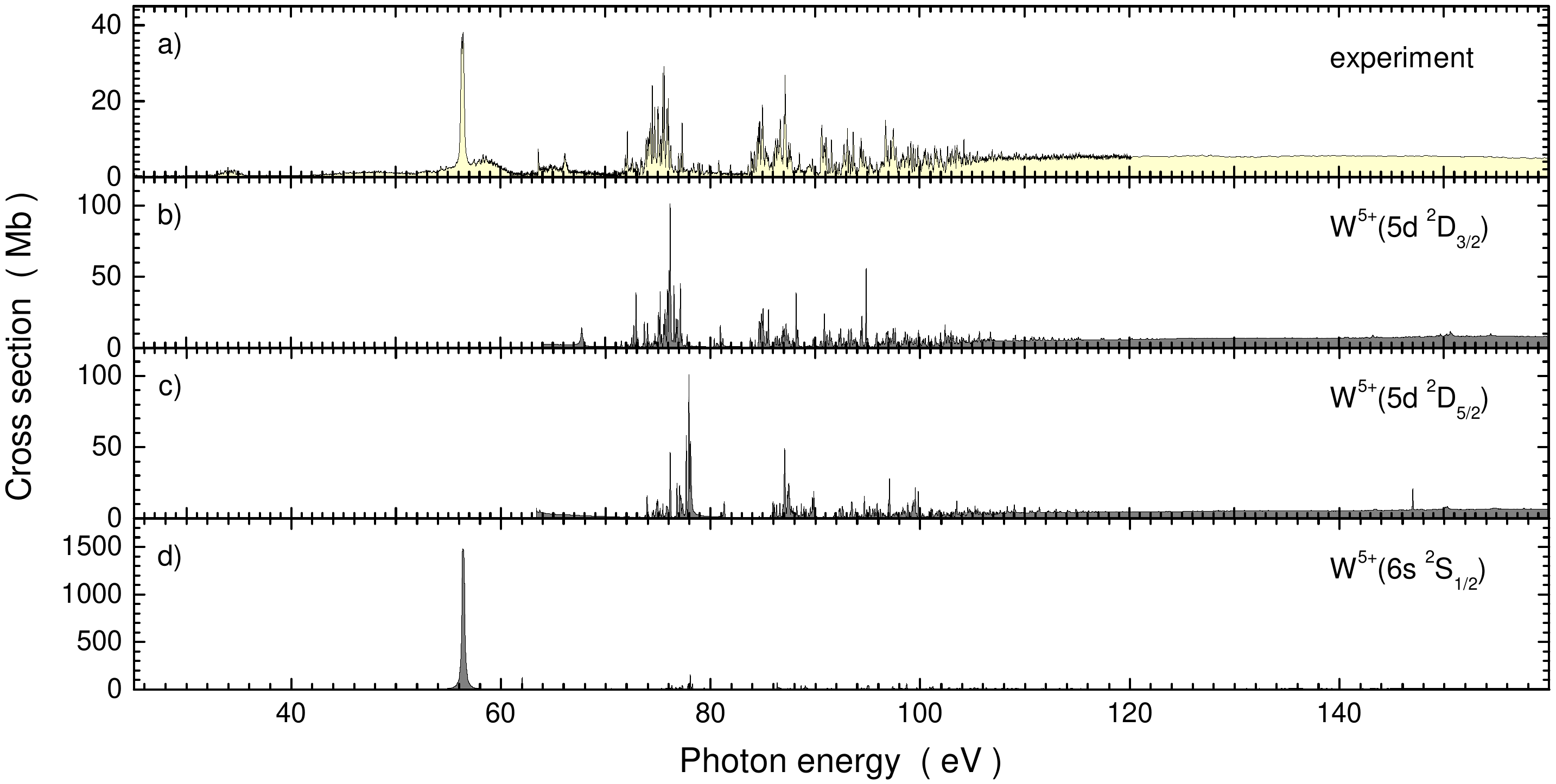}
\caption{\label{Fig:overview}(Colour online) Overview of the present experimental (panel a) and theoretical (panels b, c and d) photoionization cross sections for W$^{5+}$ ions as a function of photon energy. The experimental cross sections are shown by a solid line with light (yellow) shading. The theoretical cross sections for the three lowest even-parity levels of W$^{5+}$ are presented as the solid lines with gray shading with each panel showing the spectroscopic notation of the associated parent ion level. The calculated cross sections were obtained from the 457-level approximation and were convoluted with 67-meV FWHM Gaussians to model the experimental energy resolution.}
\end{figure*}

An overview of the experimental and theoretical results for single photoionization of W$^{5+}$ is presented in Fig.~\ref{Fig:overview}. The experimental cross section in panel a is represented by the measured energy-scan results normalized to a number of absolute measurements (not shown). The spectrum features numerous significant contributions of strong narrow resonances. Thus the character of the W$^{5+}$ photoionization cross section is strikingly different from all of the results obtained along the tungsten isonuclear sequence for W$^{q+}$ with $q = 0, 1, 2, 3$, and $4$ which are all characterized by broad cross section features, showing a transition from moderately structured spectra to mostly smooth continua as the charge $q$ decreases~\cite{Ballance2015a,Mueller2015h,McLaughlin2016a,Mueller2017b}. The experimental data were measured at a constant energy resolution which was not \textit{a priori} known. Fits to the narrowest resonances in the spectrum suggest an experimental energy spread of ($67 \pm 10$)~meV. The data were taken during a number of beamtimes at the Advanced Light Source within a time span of two and a half years. At least parts of the energy scans of the fine cross-section structures were repeated several times over the years and provided identical results. This is noteworthy because it highlights the reproducibility of ion-source performance and of the output of W$^{5+}$ ion beams with a certain composition of ground-level and metastable states.

%##########################################################################################
%
%       Fig 1
%
%##########################################################################################
%

The theoretical cross sections for W$^{5+}(5p^6 5d~^2{\rm D}_{3/2})$, W$^{5+}(5p^6 5d~^2{\rm D}_{5/2})$ and W$^{5+}(5p^6 6s~^2{\rm S}_{1/2})$ are shown in Fig.~\ref{Fig:overview} panels b, c, and d, respectively. The theoretical spectra were convoluted with 67-meV FWHM Gaussians to simulate the experimental photon energy resolution. When comparing the experimental and the theoretical photoionization spectra, the strongest resonance in the experiment at near 56.4~eV is well matched in position and shape by the dominant resonance in the photoionization spectrum of W$^{5+}(5p^6 6s~^2{\rm S}_{1/2})$. The peak heights are very different, though, with only about 40~Mb in the experiment and about 1500~Mb in the theoretical result. Although the calculated spectra for W$^{5+}(5p^6 5d~^2{\rm D}_{3/2})$ and W$^{5+}(5p^6 5d~^2{\rm D}_{5/2})$ parent ions show resonance groups at energies similar to measured dense populations of resonances, a one-to-one mapping of theoretical and experimental cross-section features is not possible.

\begin{figure}
\centering
\includegraphics[width=14 cm]{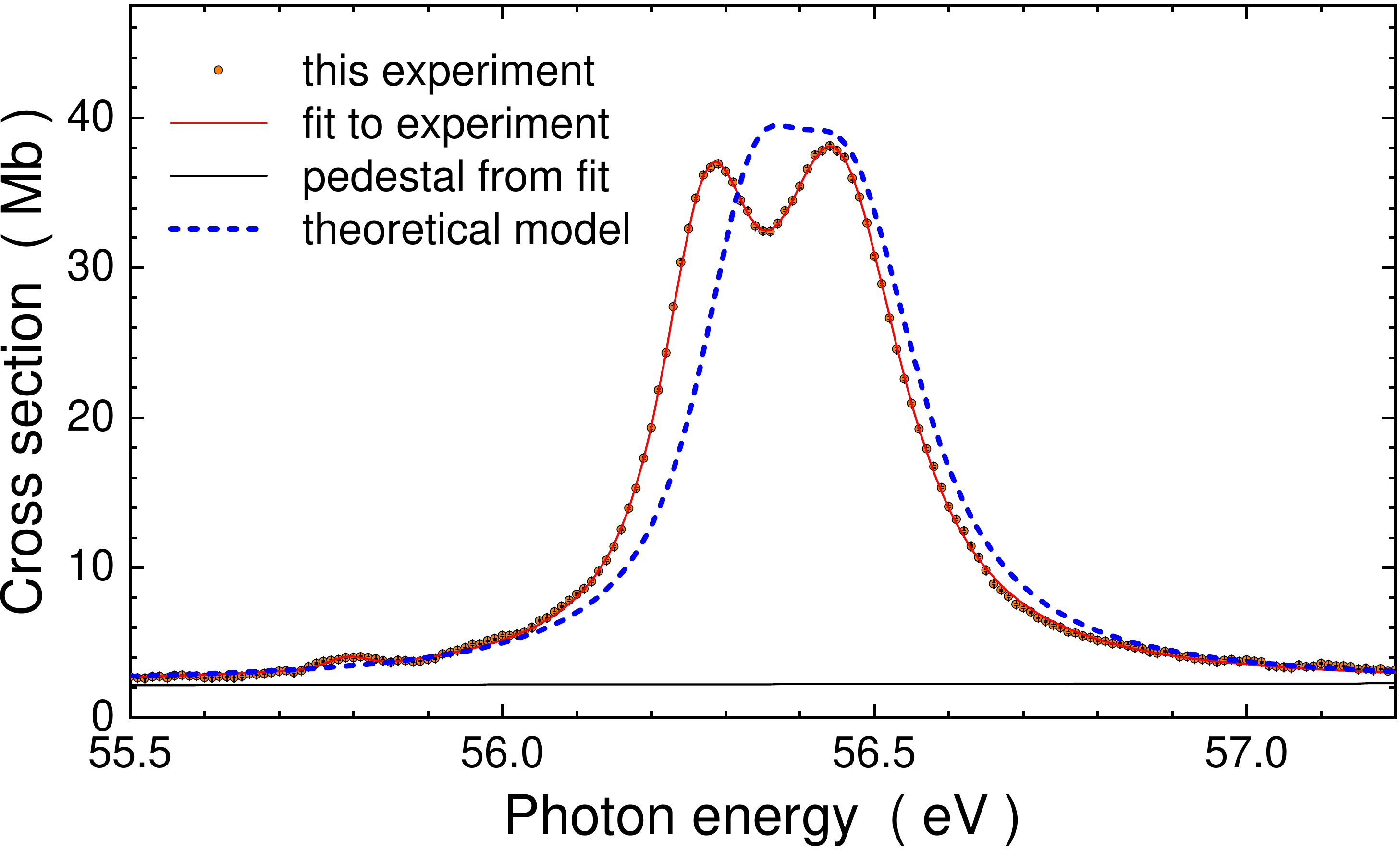}
\caption{\label{Fig:6speak}(Colour online)
Detail of the experimental photoionization measurement (solid dots) for W$^{5+}$ ions  showing the strongest resonance feature in the spectrum. Statistical error bars are smaller than the size of the symbols. The dashed line is the scaled result of the present DARC calculation for  W$^{5+}$ in the $5p^6 6s~^2{\rm S}_{1/2}$ excited level. The scaling factor is 0.025, suggesting a fraction of 2.5\% of W$^{5+}(5p^6 6s~^2{\rm S}_{1/2})$ in the parent ion beam. The solid (red) line shows the result of a three-resonances-plus-background fit to the experimental data. The horizontal solid line represents the fitted continuum cross section.
}
\end{figure}

To explore the similarity of the experimental and theoretical peak feature at about 56.4~eV, Fig.~\ref{Fig:6speak} zooms in on the photon energy range 55.0 to 57.2~eV. The experimentally observed peak clearly shows a double structure with a few small ``wiggles'' in the tails that may be associated with further small resonances. The theoretical peak was scaled down by a factor of 0.025 so that the areas under the experimental and theoretical resonances match. Apart from the smaller energy splitting in the theoretical double-peak feature, there is remarkable agreement between the measured and calculated results. The factor 0.025 can readily be explained by a fraction of 2.5\% of W$^{5+}(5p^6 6s~^2{\rm S}_{1/2})$ ions in the parent ion beam that was used in the measurements. On the basis of Cowan-code calculations~\cite{Cowan1981}, we suggest that the observed double peak is due to photoexcitation of W$^{5+}(5p^6 6s~^2{\rm S}_{1/2})$ $\to$ W$^{5+}(5p^5 6s^2~^2{\rm P}_{1/2,3/2})$, for which a calculated excitation energy of 56.93~eV is found, in good agreement with the measured value of 56.4~eV.

%##########################################################################################
%
%       Fig 2
%
%##########################################################################################
%

%##############################
%
% Table 2
%
%################################
\begin{table*}
\caption{Experimental and theoretical resonance parameters for the fine-structure levels of the W$^{5+}(5p^5 6s^2~^2{\rm P})$ term. $E_{res}$ denotes resonance energy, $\Gamma$  total width, and $\bar{\sigma}$ resonance strength. Uncertainties of the experimental quantities in the last digits are given by the numbers in brackets. They only refer to the uncertainties of the fit. On an absolute scale the resonance energies can deviate by 0.2~eV, the resonance strengths by 19\%. The difference between experimental and theoretical resonance strengths is due to the small fraction of only 2.5\% of the ions in the parent ion beam residing in the $5p^6 6s~^2{\rm S}$ level.  }
  \label{tab:params}
  \centering
  \begin{tabular}{lcccccc}
  \br
%  \hline
 % \hline
  &\multicolumn{3}{c}{$^2{\rm P}_{1/2}$}&\multicolumn{3}{c}{$^2{\rm P}_{3/2}$}\\
   &
  \multicolumn{1}{c}{$E_{res}$} &\multicolumn{1}{c}{$\Gamma$}&\multicolumn{1}{c}{$\bar{\sigma}$} &\multicolumn{1}{c}{$E_{res}$} &\multicolumn{1}{c}{$\Gamma$}&\multicolumn{1}{c}{$\bar{\sigma}$}\\ \hline
    exp. &56.275(3) &0.119(2) &5.47(12) &56.448(1) & 0.200(4) & 10.78(16)\\
    DARC &56.334 &0.142 &226.3 &56.466 & 0.215 & 420.1\\
    \mr
  %  \hline
  \end{tabular}
\end{table*}

For extracting the resonance parameters a fit was applied to the experimental data. Except for one, the ``wiggles'' in the tails of the double-peak structure were neglected in the fit. Thus, the fit function comprised three Voigt profiles superimposed upon a linear continuum function. The solid (red) line through the data points in Fig.~\ref{Fig:6speak} is the resulting fit curve. The extracted parameters are listed in Table~\ref{tab:params}.

For further comparison of theory and experiment the composition of the parent W$^{5+}$ ion beam has to be assessed. A beam containing levels $k$ with fractions $f_k$ gives rise to an apparent cross section
\begin{equation}
\sigma^{app} = \sum_k f_k \sigma_k
\label{eq:sumk}
\end {equation}
with the cross sections $\sigma_k$ for the contributing levels $k$ and with $\sum_k f_k = 1$. If theory provides good approximations $\sigma_k^{theo}$ for $\sigma_k$, the sum in Eq.~\ref{eq:sumk} should agree with the experimentally determined apparent cross section  $\sigma^{app}$ with appropriate fractions $f_k$. In the present case  the fraction $f_3$ of ions in the level $5p^6 6s ~^2{\textrm S}_{1/2}$ was estimated by comparison of experiment and theory as discussed above. For the $5d~^2{\textrm D}$ ground term a similar assessment of fractions $f_1$ and $f_2$ of the two fine-structure levels appears to be impossible on the basis of the data shown in Fig.~\ref{Fig:overview}. However, as mentioned above, fractions of different levels in a W$^{5+}$ ion beam produced with an identical type of ECR ion source were inferred by comparison of experimental and detailed theoretical calculations of apparent cross sections for electron-impact single ionization of W$^{5+}$~\cite{Jonauskas2019a}. Under the assumption that levels within a given configuration are populated according to their statistical weights, a fraction of ($85 \pm 9$)\% of ions in the ground configuration was present. With the statistical weights 4/10 of the $^2{\textrm D}_{3/2}$ level and 6/10 of the $^2{\textrm D}_{5/2}$ level, the fractions for these two levels are $f_1 = 0.34$ and $f_2 = 0.51$.

\begin{figure*}
\centering
\includegraphics[width=17cm]{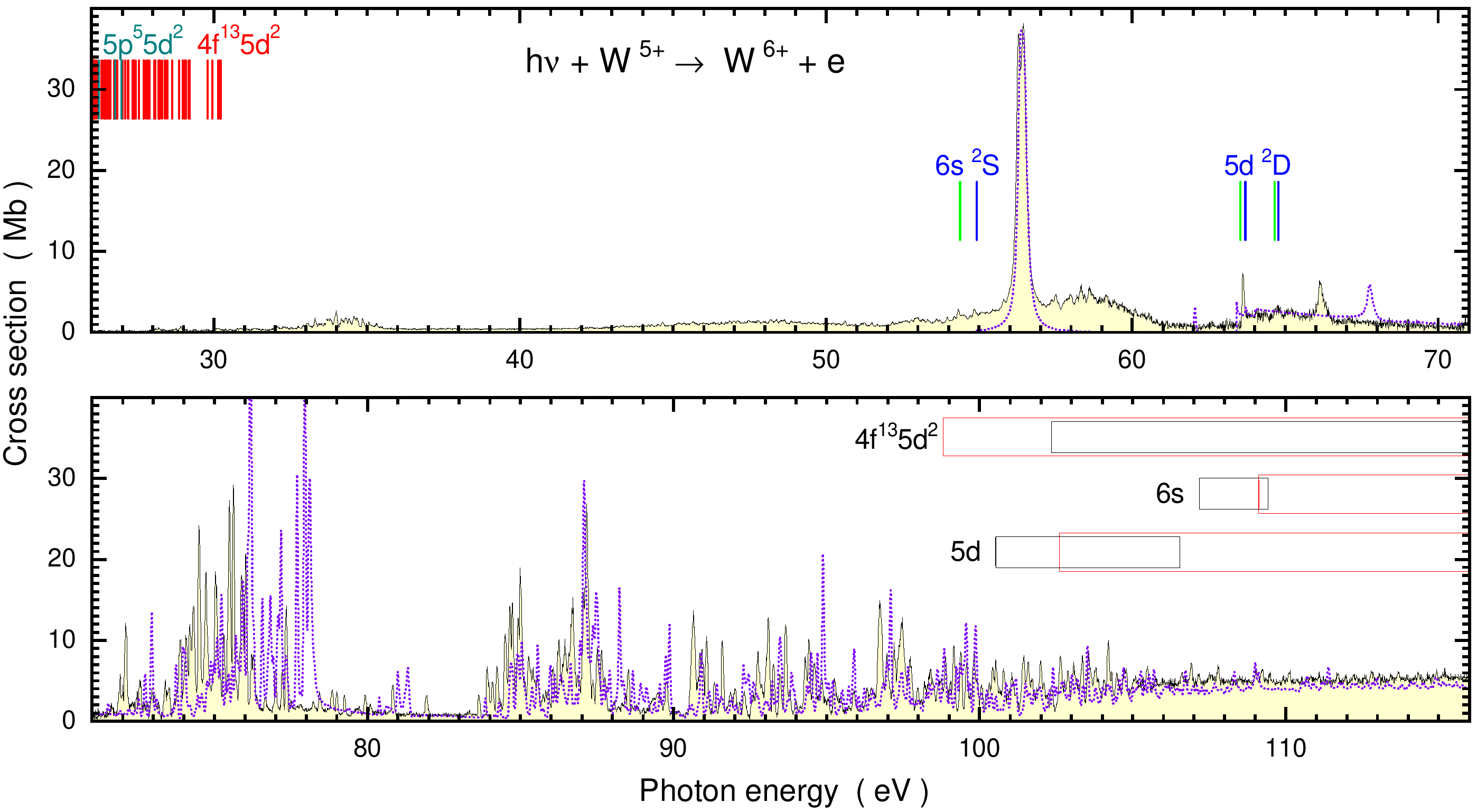}
\caption{\label{Fig:expandmodel}(Colour online)
Experimental (solid black line) and incomplete theoretical (violet dotted line, see Eq.~\ref{eq:sumDARC}) photoionization cross sections for W$^{5+}$ ions at ($67 \pm 10$)~meV energy resolution are shown in the energy range 26 - 116~eV where resonant contributions dominate the spectrum. Characteristic energies and ranges of energies are indicated: The dark (blue) vertical lines in the upper panel show the ionization thresholds of the $5d~^2{\rm D}_{3/2}$ ground level and the first two excited levels, $5d~^2{\rm D}_{5/2}$ and $6s~^2{\rm S}_{1/2}$, of W$^{5+}$ obtained from the excitation energies and the ground-state ionization threshold of W$^{5+}$ listed in the NIST Spectroscopic Database~\cite{NIST2019}. The lighter (green) vertical bars show the same thresholds calculated on the basis of the Cowan code~\cite{Cowan1981}. Also calculated with the Cowan code are the (red and dark cyan) vertical bars in the upper left corner of the upper panel. They show ionization thresholds associated with the levels in the $4f^{13} 5s^2 5p^6 5d^2$ and the $4f^{14} 5s^2 5p^5 5d^2$ excited configurations that fall within the present energy range. The boxes in the lower panel show six ranges of direct inner-shell photoionization thresholds for different configurations of W$^{5+}$ ions. The configuration-average ionization thresholds for removal of one $5p$ or $4f$ electron from three parent configurations were calculated employing the Cowan code. The two parent configurations identified by the subshell of the outermost valence electron ($5d$, and $6s$) have the core configuration $4f^{14} 5s^2 5p^6$. The parent configuration described by $4f^{13} 5d^2$ has closed $5s$ and $5p$ subshells. Thresholds for direct $5p$ subshell photoionization are within the (black-framed) boxes that have a smaller height. The (red-framed) boxes containing the thresholds for direct $4f$ subshell photoionization have a slightly increased height.
}
\end{figure*}

Since the present DARC calculations are limited to contributions of W$^{5+}$ ions in levels $5d~^2{\textrm D}_{3/2}$ (fraction $f_1 = 0.34$, calculated cross section $\sigma_1^{\small{\rm{DARC}}}$), $5d~^2{\textrm D}_{5/2}$ (fraction $f_2 = 0.51$, calculated cross section $\sigma_2^{\small{\rm{DARC}}}$), and $6s~^2{\textrm S}_{1/2}$ (fraction $f_3 = 0.025$, calculated cross section $\sigma_3^{\small{\rm{DARC}}}$), the contributions of the remaining 12.5\% of the ions present in the parent  W$^{5+}$ are not accounted for. Hence, the incomplete sum
\begin{equation}
\sigma^{theo} = 0.34 \sigma_1^{\small{\rm{DARC}}} + 0.51 \sigma_2^{\small{\rm{DARC}}} + 0.025 \sigma_3^{\small\rm{DARC}}
\label{eq:sumDARC}
\end {equation}
is compared in Fig.~\ref{Fig:expandmodel} with the experimental photoionization data.

%##########################################################################################
%
%       Fig 3
%
%##########################################################################################
%

%##########################################################################################
%
%       Fig 4
%
%##########################################################################################
%
%
\begin{figure}
\centering
\includegraphics[width=14cm]{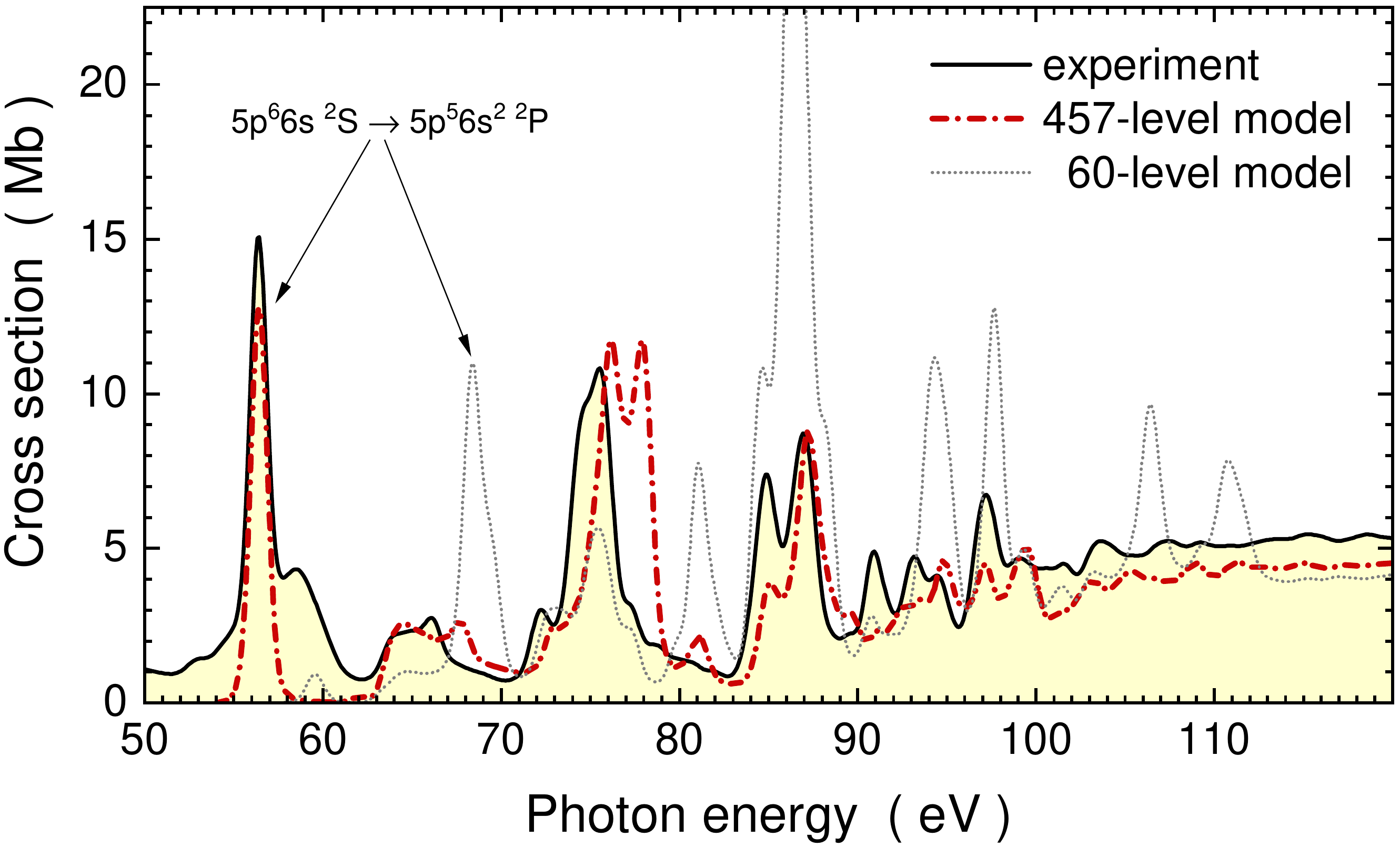}
\caption{\label{Fig:lorescomp}(Colour online)
Comparison of the present experimental and theoretical photoionization cross sections after convolution with 1-eV FWHM Gaussians. The (red) dash-dotted line represents the 457-level DARC model. The (gray) dotted line shows the result of the 60-level model.
}
\end{figure}

Obviously, contributions to the measured cross section are missed by the theoretical model. Naturally this is true for all photon energies below the $6s~^2{\textrm S}_{1/2}$ ionization threshold at about 55~eV. A broad hump found in the experiment with an onset at about 52~eV reaches up to 62~eV and is not reproduced by the present calculations. It is most likely associated with excitations from the metastable levels in configurations $5p^5 5d^2$ and $4f^{13} 5d^2$. Clearly, there is also part of the cross section above the $4f$ and $5p$ ionization thresholds missing in the theoretical cross section. The differences around 110~eV are about 15 to 20\% which can be rationalized by the missing 12.5\% of parent W$^{5+}$ ions in the theory. Assuming identical cross sections for the ionization of inner-shell electrons for ions with different valence-shell configurations, a 12.5\% reduction of the experimental cross section should match the theory model. The remaining small difference is very well within the experimental systematic-error bars of 19\%.

With the densely spaced resonances in the photon-energy range 72 to 105~eV, which are most likely due to excitations $5p \to n\ell$ and $4f \to n\ell'$, it is difficult to find unambiguous matches between theory and experiment. Clearly, the sizes of the cross sections are similar in both data sets while resonance positions do not appear to match well. The reason has already been discussed in the context of subsection~\ref{subsec:structure}, where deviations of level energies by up to 2~eV are discussed. Nevertheless, apart from an obvious energy shift, quite good agreement of theory and experiment can be recognized in the photon-energy range 63 to 69~eV. The broad resonance peak observed in the experiment at about 66.1~eV is particularly well reproduced, though shifted, in the theoretical model.

Comparison of theory and experiment is obscured by the numerous resonance features present in both spectra. %As discussed in subsection~\ref{subsec:structure}, characteristic energies in the theoretical structure calculations deviate from accepted values by as much as 2~eV, so one cannot expect that individual resonance positions in the photoionization spectrum are reproduced better than that.
To remove some of the  nontransparency introduced by the many details in the measured and calculated spectra, it is instructive to examine the data ``from a distance'', that is, to make the comparison after washing out the fine details by a convolution of both theory and experiment with a Gaussian energy resolution of 1~eV FWHM. The  theoretical spectra resulting from the 457-level and 60-level models are shown in Fig.~\ref{Fig:lorescomp} together with the convoluted experimental spectrum. The 457-level model reproduces the main features of the experimental cross section rather well. The size of the cross section and the overall structure are adequately reproduced. In contrast, the 60-level DARC model shows serious disagreement with the experiment. For example, the resonance associated with the transition ${\textrm W}^{5+}(5p^6 6s\,^2{\textrm S}) \to {\textrm W}^{5+}(5p^5 6s^2\,^2{\textrm P})$ is well reproduced by the 457-level calculation, but strongly shifted in energy when the 60-level approximation is applied. While the relative sizes of the experimentally observed resonances are not well reproduced by the 60-level model, the  continuum cross section at energies beyond about 105~eV calculated in the same approximation approximately matches the experiment. Clearly, the calculations employing the much larger 457-level basis set are superior to the 60-level model. It should be noted in this context that as discussed above, contributions from 12.5\% of the parent ions  are not accounted for in the theory.

\section{Summary and Conclusions}\label{sec:Conclusions}
Experimental and theoretical photoionization cross sections for  W$^{5+}$ ions are presented.
The experimental cross sections were measured on an absolute scale employing the photon-ion merged-beams technique. The theoretical data were obtained from large-scale close-coupling calculations
within the Dirac-Coulomb R-matrix approximation (DARC).  Comparison of the measured and calculated results for W$^{5+}$ is complicated by the presence of many long-lived excited states in the parent ion beams used for the experiments. The true difficulty, however, is in the calculation of the exact electronic structure of the W$^{5+}$ ion with its high degree of complexity, particularly in the many excited autoionizing levels.

The DARC calculations for W$^{q+}$ with $q = 0, 1, 2, 3$, and $4$ show increasingly better agreement with experiments along the sequence of increasing charge states. This was expected because the physics of more highly charged ions becomes simpler in that the electron-nucleus interactions become more prominent relative to the electron-electron interactions that are difficult to treat.  In photoionization of W$^{5+}$ the sequence is disrupted, as the character of the experimental cross section changes from a few broad peak features to a multitude of narrow resonances. The sophisticated DARC calculations provide only semi-quantitative predictions for this case. At first sight, this is surprising, since the ground-state configuration of W$^{5+}$ is characterized by a single $5d$ electron outside closed electron shells, that is, ground-term W$^{5+}$ is a quasi--one-electron system. However, the photoexcited levels feature two or mostly even three open subshells with vacancies in the $5p$ or the $4f$ subshells. This makes the exact calculation of individual narrow resonances almost impossible. Hence, while broad photoionization features in W$^{q+}$ ions with $q\leq4$ are forgiving of small deviations in energy calculations, the narrow-resonance photoionization spectrum of W$^{5+}$ clearly reveals the limitations of calculations for such complex ions, even when hundreds of basis states are considered.

\ack

AM acknowledges support by Deutsche Forschungsgemeinschaft
under project numbers Mu-1068/20 and Mu-1068/22. RAP acknowledges support from the US Department of Energy (DOE) under grant No. DE-FG02-03ER15424.
BMMcL acknowledges support by the US National Science Foundation through a grant to ITAMP at the Harvard-Smithsonian Center for Astrophysics, the award of an adjunct professorship from the University of Georgia at Athens,  and Queen's University Belfast
for a visiting research fellowship (VRF).
The computational work was carried out at the National Energy Research Scientific
Computing Center in Berkeley, CA, USA and at the High Performance
Computing Center Stuttgart (HLRS) of the University of Stuttgart, Stuttgart, Germany.
The authors gratefully acknowledge the Gauss Centre for Supercomputing e.V. (www.gauss-centre.eu) for funding this project by providing computing time on the GCS Supercomputer
HAZEL HEN at H\"ochstleistungsrechenzentrum Stuttgart (www.hlrs.de).
The Advanced Light Source  is supported by the Director, Office of Science, Office of Basic Energy Sciences,
of the US Department of Energy under Contract No. DE-AC02-05CH11231.
%\clearpage
%+++++++++++++++++++++++++++++++++++++++++++++++++++++++++++++++++++++++++++++
%
%   Reference section now follows
%
%   Delete or change fake bibitem. delete next three
%   lines and directly read in your .bbl file if you use bibtex.
%
%+++++++++++++++++++++++++++++++++++++++++++++++++++++++++++++++++++++++++++++
%
\section*{References}
\bibliographystyle{iopart-num}
%\bibliographystyle{unsrt}
%\bibliography{wions}
%\bibliography{k3AM}

\providecommand{\newblock}{}

\end{document}